# A Rough Sets Partitioning Model for Mining Sequential Patterns with Time Constraint

Jigyasa Bisaria[1], Namita Shrivastava[2] and K.R. Pardasani[3]
Department of Mathematics
Maulana Azad National Institute of Technology (A Deemed University)
Bhopal 462051 India
[1]jigyasa@bisaria.com , [2]sri.namita@gmail.com, [3]kamalrajp@hotmail.com

*Abstract*— now a days, data mining and knowledge discovery methods are applied to a variety of enterprise and engineering disciplines to uncover interesting patterns from databases. The study of Sequential patterns is an important data mining problem due to its wide applications to real world time dependent databases. Sequential patterns are inter-event patterns ordered over a time-period associated with specific objects under study. Analysis and discovery of frequent sequential patterns over a predetermined time-period are interesting datamining results, and can aid in decision support in many enterprise applications. The problem of sequential pattern mining poses computational challenges as "a long frequent sequence" contains enormous number of frequent subsequences.Also useful results depend on the right choice of event window. In this paper, we have studied the problem of sequential pattern mining through two perspectives, one the computational aspect of the problem and the other is incorporation and adjustability of time constraint. We have used Indiscernibility relation from theory of rough sets to partition the search space of sequential patterns and have proposed a novel algorithm that allows previsualization of patterns and allows adjustment of time constraint prior to execution of mining task.The algorithm Rough Set Partitioning is atleast ten times faster than the naive time constraint based sequential pattern mining algorithm GSP. Besides this an additional knowledge of time interval of sequential patterns is also determined with the method.

## I. INTRODUCTION

Data mining is a technique that uncovers useful patterns hidden in various real world databases. A lot of research effort on techniques of datamining can be cited in recent literature. One of the key research areas in datamining is "mining frequent sequential patterns in large databases".

Sequential pattern mining finds inter-event patterns ordered over a period of time and associated with an object under study. The study of frequent sequential patterns gives useful predictive knowledge about patterns of "antecedent is followed by consequent" form. These event patterns are inherent in many real world databases and hence the concept is widely applicable to a variety of analytic disciplines. It is applied to study root causes of banking customer churn [7], analysis of telecom alarm sequences [8], analysis of web browsing patterns [9] & study of adverse drug reactions as temporal association rules [10]. The problem of sequential pattern mining in databases was introduced by Agrawal and Srikant [1] and Manilla et al. [2] at about the same point of time. Agrawal and Srikant [1] coined the concept based on famous example of market basket analysis. They defined an itemset as "a set of items purchased by a customer" in a transaction. If we group the records of the itemsets purchased by the same customer in different transactions, we derive the sequence of items purchased by the customer.

Thus, an itemset is a non empty set of items and a sequence is a non empty set of itemsets. The size of a sequence is number of events in it. A pattern is interesting if its frequency of occurance in the database crosses a user specified minimum numeric value called the support threshold.

Consider an example of a customer record in the transaction database of a retail store, suppose a customer purchases a computer on 1/1/2008 and returns to buy a printer on 15/2/2008. Again the same customer purchases a pen drive on subsequent visit on 19/3/2008. The sequence of purchase pattern for this customer will be computer→printer→pen drive. Suppose now the same customer returns on 15/3/2009 for buying a new computer. If we study the sequence of events of this cutomer without any presumption of time interval the new sequence corresponding to this customer would be computer→printer→pen drive→computer.

In most of real world problems, especially pattern study for managerial decision support, it is important to impose time interval constraint in the sequential pattern mining task. Let us take an example of another application; that is the study of fault patterns from a specific location in telecom landline networks. The sequential pattern under study is derived by the choice of appropriate time interval. Table I represent a sample information system that collects fault signals from telecom landline networks and Table II and Table III represent the resultant sequence databases formed by chosing appropriate event window. It is evident that the length of a sequence depends on the choice of event window under consideration. Also the quality of mining results depends on the appropriate choice of event window. Consider a pattern A followed by B over a period of 3 months and same pattern within a time interval of 1 year, both patterns are not only dissimilar but can result in different inferences. While the former may indicate an association between events, the latter may not be an implication relationship. Also additional knowledge about time interval of frequent patterns can lead to significant additional mining information.

**TABLE I Information System**

| Location ID | Signal date | Signal |
|---|---|---|
| 1 | 10/5/2008 | 10 |
| 1 | 12/5/2008 | 20 |
| 1 | 15/5/2008 | 30 |
| 1 | 16/5/2008 | 50 |
| 1 | 25/5/2008 | 40 |
| 2 | 15/5/2008 | 20 |
| 2 | 23/5/2008 | 40 |
| 3 | 12/5/2008 | 10 |
| 3 | 17/5/2008 | 50 |
| 3 | 20/5/2008 | 60 |
| 3 | 25/5/2008 | 70 |
| 4 | 15/5/2008 | 10 |
| 4 | 23/5/2008 | 70 |
| 4 | 25/5/2008 | 60 |

**TABLE II Derived Sequence table time interval $10^{th}$–$25^{th}$ May**

| Location ID | Time Interval | Sequence of events |
|---|---|---|
| 1 | 15 | 10:20:30:50:40 |
| 2 | 15 | 20:40 |
| 3 | 15 | 10:50:60:70 |
| 4 | 15 | 10:70:60 |

**TABLE III Derived Sequence table time interval $15^{th}$–$25^{th}$ May**

| Location ID | Time Interval | Sequence of events |
|---|---|---|
| 1 | 10 | 30:50:40 |
| 2 | 10 | 20:40 |
| 3 | 10 | 50:60:70 |
| 4 | 10 | 10:70:60 |

There are two aspects in the study of sequential pattern mining; one aspect is the computational methodology as the problem complexity is large and the other aspect is usability of the method and incorporation of time constraint in the mining task.

Various studies on the computational aspect of the problem can be cited [1]-[6]. Very few of these methods incorporate the time constraint in the mining task. Incorpation of time constraint is important in for generation of quality mining results.

A variation of sequential pattern mining as sequential pattern mining with constraints [14]-[17] can help in adjustment of time interval. Any user of the algorithms pre-sets the time interval and then runs the mining task. If he/she chooses a short interval, the emergent pattern may lack useful predictive knowledge. However, a long interval chosen may generate huge number of irrelevent patterns. The resultant patterns emerge only after completion of full mining task. Thus, every adjustment of event interval requires the mining task to be reiterated.

In this paper, we have addressed both the issues related to present methods for mining sequential patterns. (i) We have proposed a user friendly interface that generates previsualization of a sample of emerging sequential patterns and allows flexible imposition of time constraint prior to mining task and (ii) we have presented a novel algorithm based on indiscerniblity relation from theory of rough sets to address the computational aspect of the expensive mining problem of frequent sequential patterns. It is found from experimental evaluations that our algorithm is atleast 10 times faster than algorithm GSP [3].

## II. RELATED WORK

Various computational methods for mining sequential patterns have been proposed by various researchers ever since the concept was coined simultaneously by Agrawal and Srikant [1] and Mannila et al. [2]. A sequence is frequent if it crosses a user specified minimum support threshold. Agrawal and Srikant [1] proposed a series of approaches for mining frequent sequential patterns based on apriori heuristic, which states, "Any subset of a frequent itemset is frequent". The best method Apriori ALL is a three-phase algorithm: (i) find frequent itemsets, (ii) create transformed database (a database formed by removing infrequent itemsets) and (iii) mine frequent sequences. The method had the following drawbacks: lack of time window constraint and expensive tranformation phase. Moreover, it is impossible to adjust the time window with the proposed method.

Mannila et al. [2] introduced the concept of sequential pattern mining as "frequent episodes". They defined "an episode as a collection of events that occur relatively close to each other in a given partial order." They did consider the importance of time frame of patterns and gave the concept of event window and sliding event window. They defined patterns as directed acyclic graphs with vertex as a single event and edge as "Event A occurs before event B". Their method of finding frequent episodes is "bottom-up candidate-generate and test apporach" which is similar to Apriori ALL proposed by Agrawal and Srikant [1].

Agrawal and Srikant [3] improved their work on mining sequential patterns in [1] and incorporated maximum gap, minimum gap and user defined taxonomies in the mining task. The new method GSP (Generalized Sequential patterns) is about 20 times faster than Apriori ALL. They introduced a bottom-up approach to search frequent patterns involving multiple scans of the database under study. The method first finds occurance frequency for all length-1 sequences that satisfy the user constraint of maximum & minimum time gap and also cross the minimum support threshold. This becomes the seed set for further iteration. The Candidate Length-2 sequences are formed by joining the elements of the seed set. Now, the database is scanned again for searching these candidates and their counts are accumulated after checking the time gap constraint. In subsequent iterations, candidate k-length sequences are formed by joining frequent k-1 sequences that have the same contiguous subsequences. Suppose a sequence $S_\alpha = \langle e_1, e_2, \ldots e_n \rangle$, another sequence $s_\beta$ is a contigeous subsequence of $S_\alpha$ if (i) $s_\beta$ is derived from $S_\alpha$ (ii) $s_\beta$ is derived from $S_\alpha$ by dropping an item from an element $e_j$ that has at least 2 items. (iii) $s_\beta$ is a contiguous subsequence of $s_\delta$ and $s_\delta$ is a contiguous subsequence of $S_\alpha$

The process is continued untill all frequent sequences present in the database are found. GSP is a multiple scan method based on apriori. Since the time-window constraint is build into the mining process the user will have to rerun the algorithm for adjusting the event window.

Jay Arey et al. [4] introduced SPAM Sequential Pattern Mining using a Bitmap representation. SPAM performs well

on databases with long sequential patterns. The algorithm is a Depth First Search method in the tree of lexicographically arranged sequences. It is candidate-generate and test method with pruning mechanism based on apriori heuristic. For efficient counting database is transformed into a bitmap format in a way that if the $i^{th}$ item is found in $j^{th}$ transaction the bitmap corresponding to that item is set to 1 otherwise 0. SPAM does not incorporate any time window constraint and hence there is no flexiblility of interval adjustment. In addition, memory consumption is high the algorithm generates even those candidates that do not occur anywhere in the database under study.

M. Zaki [5] proposed a lattice-based approach that uses a vertical id representation of data. SPADE (Sequential patern discovery using equivalence classes) breaks down the problem into smaller subproblems and uses simple join operations for the mining task. It requires maximum 3 database scans. The method uses a vertical id list representation of database. The data is grouped based on the same prefix and a lattice of equivalence classes is formed. Then two methods Breadth First Search and Depth First Search are applied on the lattice of equivalence classes.The baseline method is again apriori like candidate-generate and test method. Due to partitioning of search space, it performs better than ApioriALL [1] and GSP [3].

All the above methods in [1]-[5] are candidate-generate and test methods.Many of these candidates genertred may not have a single occurance in the database under study. Pei et al. [6] proposed an algorithm for mining sequential patterns named PrefixSpan with Pseudoprojection. This method studies only those sequences that exist in the database under study. The algorithm is a database projection scheme wherein a sequence database is recursively projected in a set of smaller projected databases based on current sequential patterns. Sequences grow on the basis locally frequent fragments. The method is very efficient for small datasets. It outperforms GSP [3] & SPADE [5] but looses in efficiency to SPAM [4] for many large databases. The method neither talks about time window constraint nor allows the flexibility of the same and thus may generate enormous useless patterns.

Most of the efficient algorithms for sequential patterns mining do not incorporate time constraint or if they do so, they do not allow the adjustment of the same prior to execution of the mining task. Thus, any adjustment in the event window will cause a rerun of the computationally expensive mining task.

A variation of sequential pattern mining as *sequential pattern mining with constraints* [14] [15] can help in adjustment of time interval. Any user of the algorithms pre-sets the time interval and then runs the mining task. If he/she chooses a short interval, the emergent pattern may lack useful predictive knowledge. However, a long interval chosen may generate huge number of irrelevent patterns. Even an experienced analyst will have to rerun the computationally intensive mining task as he/she can view the emerging patterns only after execution of full mining task. Chen et.al [16] have modified aprioriAll [1] and prefixspan [5] to generate additional knowledge of time interval of sequential patterns. Hirate and Yamana[17] have emphasized on the imposition of time constraint in the mining of sequential patterns and have presented a variation of prefixspan[5] method for mining of time constrained sequential patterns. Again the method in [16] merely generates addition information of time interval of patterns and the method proposed in [17] has the same limitations as methods proposed in [14] [15].

We identify the following three issues from above litrerature survey and focus on these issues in our proposed algorithm. (i) Most of the methods being candidate generate and test methods; lot of computation time is spent in candidate generation and testing. (ii) How to incorporate time constraint while still keeping the mining task economical? (iii) Flexibilty of event window adjustment by providing scope for human intervention for generation of only relevent patterns. This motivated us to design a user-friendly algorithm that seperates the choice of event window from the mining task by providing the user a pre-visualization of patterns. An experienced analyst can tailor the event window by viewing a sample of records from the database under study. This allows the algorithm user the flexiblity of adjustment of event window for generation of useful implication relations. If the user is not sure of the time-interval to choose, he/she can choose a time window so that all the records in the enterprise database are considered.

### III. PROPOSED MODEL AND METHOD

Let $I = \{i_1, i_2, i_3, \ldots, i_n\}$ be the set of items in a customer transaction database. In above example of database of fault signals as given in TABLE I the items correspond to fault signals from a specific location. An itemset is a set of items that occur together in a transaction at a point of time. A sequene is a set of events formed by grouping itemset associated with an object under study and are ordered over a period of time. A sequence database is a as given in TABLE II and TABLE III formed by choice of different time interval. Formally,

Definition 1: Information System [11]: FroSm theory of rough sets, an information system is given as: $S = \{U, At, V, f\}$ where $U$ is a finite set of objects, $U = \{x_1, x_2, \ldots, x_n\}$ At is a finite set of attributes, At is further classified into two disjoint subsets, conditional attributes C and decision attributes D, $At = C \cup D$  $V = \bigcup_{p \in At} V_p$ and $V_p$ is a domain of attribute p

$f : U \times At \to V$ is a total function such that $f(x_i, q) \in V_q$ for every $q \in At$ and $x_i \in U$

Example: Let us consider the information system given in Table I. The set U corresponds to data label LocationID. The set of attributes are event time stamp and fault signals.

Defintion 2: event interval (time-interval): Suppose $t_s$ is the start Time and $t_e$ is the end time for study of event patterns. Then, the event/time interval for study of patterns is given by: $t_s - t_e$ for given information system S and is derived by $T \subset At$

Definition 3: Sequence or serial episode: A sequence or a serial episode is defined as a set of events that occur within a predetermined event interval and are associated with the object under study. Let I be the set of itemsets $I = \{i_1, i_2, i_3 \ldots \ldots i_n\}$ then the set of sequence $E \subset At$ is formed by combining itemsets associated with the same object and are ordered by time $E = \{e_1, e_2, e_3, \ldots \ldots e_m\}$ each $e_i = \{i_1, i_2, \ldots \ldots i_l\}$ is some combinations of itemsets.

Definition 4: Length of a sequence: The length of a sequence is the number of items it contains. A k-sequence contains k items $k = \sum_j |e_j|$

Definition 5: Indiscernibity realtion: Let $B \subset At$ $x_i, x_j \in U$ It can be treated as a binary relation called Indiscernibility relation as: $IND = \{(x_i, x_j) \in U^2 : \forall p(x_i) = p(x_j)\}$

It can be said that $x_i, x_j$ are indiscernible by a set of attributes B in S if and only if $p(x_i) = p(x_j)$ for every $p \in B$ Clearly, IND is an equivalence relation on U. Equivalence classes of relations is called elementary sets in S. For any element $x_i \in U$ the equivalence class of $x_i$ in relation B is represented as $[x_i]_{IND}$

Definition 6: Subset and Subsequence relations: A set X of itemsets is said to be a subset of set Y if set X is contained in set Y and is denoted as $X \subseteq Y$ For example, if set $Y = \{1, 2, 3\}$ and the set $X = \{1, 2\}$ then $X \subset Y$ A sequence $Y' = \langle y_1, y_2, y_3 \ldots \ldots y_n \rangle$ is said to be a subsequence of sequence $X' = \langle x_1, x_2, x_3 \ldots \ldots x_m \rangle$ if $\exists \ i_1, i_2, i_3 \ldots i_k \in x_i$ such that $y_1 \subset x_1, y_3 \subset x_3 \ldots y_n \subset x_m$, For example, the sequence $\langle 1:2 \rangle$ is a subsequence of $\langle (1,2,4) : (2,4) \rangle$ However, $\langle 1:2 \rangle$ is not contained in $\langle 1, 2 \rangle$ since both have different interpretation while $\langle 1:2 \rangle$ are events seperated by order of time, $\langle 1, 2 \rangle$ are parallel events that is events that occur in same transaction time. For the sake of readability the serial episodes are seperated by ":"

Definition 7: Support of a sequence: The support of a sequence "a is followed by b" is defined as:

$$\frac{\text{Total customers that contain the sequence a} \rightarrow b}{\text{Total number of customers}}$$

A sequence is frequent if it crosses a user specified minimum support threshold.

Definition 8: Sequential pattern mining in constrained event window: The problem of mining sequential pattern within a time interval is to find all event sequences that cross a user specified minimum support threshold and are in the constrained time interval under consideration. The problem can be split into two parts, one mining frequent parallel episodes (items that occur together in a transaction) and second mining frequent serial episodes (events in order of

time).Consider Figure 1.0 Clearly sequence of events under time window $\langle t_s | t_e \rangle$ is $\{e_1, e_2, e_3\}$ and under event window $\langle t_s | t_{e'} \rangle$ is $\{e1, e2\}$. $t_s$ represents the start time and $t_e$ and $t_{e'}$ gives the end time.

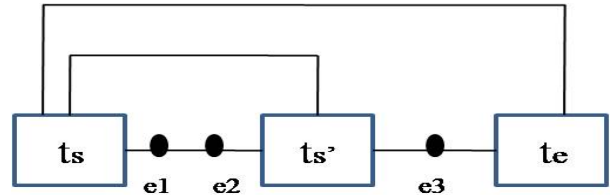

**FIG. 1.0 EVENT SEQUENCE**

It is evident that the length of a sequence depends upon the time interval under consideration. A long time interval will cause a long sequential pattern. In some applications, especially pattern study for managerial decision support, an experienced professional need to view only a sample of sequential patterns to decide the time window for generation of useful implication relations. Many times viewing a sample of sequences can motivate an adjustment in the event window under consideration until finally we find an optimum event window for generation of useful implication relations.

The proposed model employs a combination of partitioning approach with Rough sets to mine frequent sequences. It uses a divide and conquer strategy to explore the search space of frequent sequences. This Rough Sets Partitioning (RSP) model is a three-stage strategy; first step is to freeze the time interval with previsualization of a sample of records. Figure 2.0 depicts the interactive window that allows the user to query the transaction table and get a previsualization of a sample of sequences that emerge out of the choice of the event window. The event window can be adjusted accordingly so that a close to the optimum sequence length is obtained for sound predictive implication relations. This event window is freezed by an experienced analyst from the domain under study, as he/she can judge the emerging patterns through previsualization.

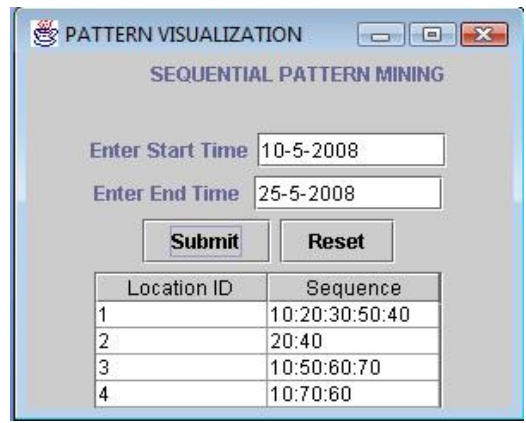

*Fig. 2.0 Sequential Pattern Previsualization*

The information System is as depicted in Table I and from the above step we derive the input to the problem of frequent sequential pattern mining which is a set of patterns which are indiscernible by the attribute time-interval. By this step we are incorporating the concept of human guided mining of patterns. This insures the discovery of useful implication relations. The algorithm pseudocode of PL/SQL that derives the sequence table from Transaction Table is as given in Figure 3.0:

```
Π Top k LocationId from Table1 where transaction_date>=Tstart &
transaction_date<=Tend
//--Π is project operator of relational algebra which implies Select
Distinct k is the number of records the user //wishes to visualize
  FOR each customer id in the rec_inner_test
       LOOP
            return_str:='';
       FOR I IN 1..rec_inner_test.COUNT
       LOOP
       return_sequence:=return_str||rec_inner_test(i).signal||':'
            END LOOP;
Update the Sequence_table with Sequence against each LocationID
ENDLOOP
```

*Fig. 3.0 Algorithm Pseudocode to derive sequences from transaction database*

We separate the time interval constraint from the mining task with the above PLSQL pseudocode and save computational effort of iteratively checking the same in the algorithm that addresses the problem computationally. The next problem is to derive sequential patterns that cross the user defined minimum support threshold from information system as in TABLE II. There are two types of patterns in a sequence database; one parallel episode (itemsets) and another set of serial episodes as given in Definition 3. Here in the depicted example under consideration the itemset is a singleton set. This is common to many applications for example database of DNA sequences, database of web logs etc.

Lemma 1: The set of all frequent sequences can be derived from frequent itemsets. [5]

Proof: Sequential patterns are inter-event patterns. Given a transaction database as in Table I, Table II of sequence is formed by grouping itemsets with the same Location ID and arranging it in order of chosen time interval. Thus the set $E \supset I$ Let $\hat{FS}$ be the space of all frequent itemsets (parallel episodes) & let $\hat{FS}'$ is the space of all frequent sequential patterns. Clearly since all sequences are derived from itemsets, any itemset with occurance less than minimum support cannot contribute more than the support threshold while accumulating counts of sequences. Hence, $\hat{FS}' \supseteq \hat{FS}$

**Method RSP:** The input to the problem of sequential pattern mining is an information system as in TABLE II having the Attribute set of sequence of events derived by finding patterns indiscernible by the attribute time-interval and associated with the object under study. From Lemma 1, we ca prune the search space of serial events by pruning out all those items which are not frequent. We scan the database once and store all the data in the attribute set of events into two datastructures. One is the domain of set E containing all unique sequences and itemset in S.

Step 1: From Definitiom 1, we consider the $E \subset At$ as the set of event sequences. We find $V = \bigcup_{e_i} V_E$, where $V_E$ is domain of attribute E.

Now to find frequent items, we query all unique itemsets and store them in a set $\hat{I}$. Now we employ indiscernibility mapping as:

$$IND(E) = \{(e_i, e_j) \in \hat{I} \times E\}$$

In this operation, we maintain an array of frequencies of the size of $\hat{I}$, we read each element of and seach for its occurance in E, and accumulate its frequency as it is found in E. We prune from V all elements of $\hat{I}$ which do not satisfy the minimum support threshold. It is clear that E & $\hat{I}$ are ways to represent event elements corresponding to U and frequency count corresponds to number of elements in U which satisfy the indiscernibility relationship. Inderscernibility mapping is a search function that finds occurance of $\hat{I}$ in set E.

Example: Consider the information system in Table II, the set $\vec{I}$={10,20,30,40,50,60,70} and set E={10:20:30:50:40,20:40, 10:50:60:70, 10:70:60} Step 1 will accumulate counts of each element in $\vec{I}$ in E. Suppose the minimum support count is 2, The Space of frequent items comprises of Set F={10,20,40,50,60,70}. From Lemma 1 it is clear that only those sequences can be frequent which are formed by elements in F.

Lemma 2: Let m denote the number of frequent items. Then the total number of sequences of length atmost k is $O(m^k)$ [5]

Proof: we will count the number of ways in which a k-sequence can be constructed and then assign items for each arrangement. For an event of length i, we have $\binom{m}{i}$ item assignments. Multiplying the choices of each case, and adding all the cases we obtain the total numbr of k sequences. Given as: $\sum_{i_1=1}^{k} \binom{m}{i_1} \sum_{i_2=1}^{k-i_1} \binom{m}{i_2} \cdots \sum_{i_3=1}^{k-i_1-i_2\cdots i_{k-1}} \binom{m}{i_k}$

The upper bound on the number of sequences of m items is $2^{k-1}, m^k = O(m^k)$, since we count both parallel and serial episodes.

Step 2: Now we have to deduce the space of frequent serial episodes. From Lemma 2, it is evident that the search space is large. We use Lemma 1 to reduce the maximum length of a serial pattern. Now we evaluate all possible serial patterns which are supersets of elements in F. This is done by breaking up the elements of into elements with its respective subsequence. The set so formed is the power set comprising of the all element the elements of domain V of E.

Illustrative Example: Consider E= {10:20:30:50:40, 20:40, 10:50:60:70, 10:70:60} the respective set P (E) is given by:
P(E)={10, 10:20, 10:40, 10:50, 10:60, 10:70, 10:20:50, 10:20:40, 10:50:60, 10:50:70, 10:70:60, 10:20:50:40,

10:50:60:70, 20, 20:40, 20:50, 20:50:40, 40, 50, 50:40, 50:60, 50:70, 60, 60:70, 70, 70:60, 50:60:70}

It is evident that our information system representation P (E) does not have a decision attribute. We create the decision attribute as the prefix of the sequence under study. And partition the database into elements with the same prefix. Thus the set D is the prefix of the elements in P(E).

Property 1: If $A \subseteq B$ for sequences A and B then Support (A) ≥ Support (B) since all transaction in information system S that supports B also necessarily also support a [12]

Illustrative example: The following partitions are created in P (E) on the basis of prefix.

| IND(10) | IND(20) | IND(40) | IND(50) | IND(60) | IND(70) |
|---------|---------|---------|---------|---------|---------|
| 10 | 20 | 40 | 50 | 60 | 70 |
| 10:20 | 20:40 | | 50:40 | 60:70 | 70:60 |
| 10:40 | 20:50 | | 50:60 | | |
| 10:50 | 20:50:40 | | 50:70 | | |
| 10:60 | | | 50:60:70 | | |
| 10:70 | | | | | |
| 10:20:40 | | | | | |
| 10:20:50 | | | | | |
| 10:50:60 | | | | | |
| 10:50:70 | | | | | |
| 10:70:60 | | | | | |
| 10:20:50:40 | | | | | |
| 10:50:60:70 | | | | | |

*Fig 4.0 Partitions on the basis of prefix indiscernibility*

Step 3: Now, we use dynamic frequency counting method in each partition to accumulate frequency of occurance of serial episodes. For this we need to maintain an array of frequencies of the size of all the elements bifurcated in partitions based on indiscernibility relation. We exclude items with cardinality 1 from this process as they are already accounted for step 1. We use property 1 for frequency accumulation for sequences with cardinality>1. The following steps explain the same:

Step 3.1: Scan E

Step 3.2: Deduce subsequences, check if the subsequece is a superset of element in F.

Step 3.3: Each element subsequence found accounts for an increment in the element frequency at appropriate index in partition and one increment to its subset, the process continues till all elements of E are considered.

Illustrative example: suppose we are considering the element 10:20:30:50:40 for frequency accumulation, since 30 is not a member of F, the sequence under consideration will reduce to 10:20:50:40. The frequency accumulation process will update the frequency of 10:20, 10:50, 10:40, 10:20:40 and 10:20:50 and lastly 10:20:40:50 in its prefix inderscernibility. Further, subsequences 20:40, 20:50 and 20:50:40 are searched in its equivalence classes and their frequencies are also updated. This step saves the computation time required to search for an element from scatch for accumulating its frequency. The process continues till all elements of E are considered. We prune elements that do not satisfy the support threshold. Also since all subsequences of a frequent sequence are frequent the search method is not strictly bottom up.

We achieve a reduction in complexity and computation by using RSP. The complexity reduction is achieved by limiting k which is maximum length of of sequence by imposing time constraint and Lemma1 as compared to the case when full dataset is considered. The computation effort is reduced by Lemma1, Lemma2 and dynamic support accumulation by using property 1. RSP works on the break and find strategy. Thus, we access only those elements as potentially frequent which have alteast a single occurance in the database under study. So, if there are q frequent itemsets and the maximum length of the sequence is less than l due to time window constraint and pruning due to lemma 1; then the upper bound of the algorithm is $O(q^l) \leq O(p^k)$. Since $q \leq p$ and $l \leq k$.

---

**Input:** Information System as in TABLE II derived by user specified time interval and mininmum support threshold
**Output:** Space of all frequent itemsets and sequences
**Method:** $E \subset At$  Find V. Derive $\hat{I} \subset V$

For each $i \in \hat{I}$ find frequency of occurance in E by
$IND(E) = \{(e_i, e_j) \in \hat{I} \times E\}$

Derive $F \subseteq \hat{I}$ //F is obtained by pruning infrequent itemsets
//Find set P(E) derived from set E and set F
For each $e_i \in E$

$e_i' = subsequence(e_i)$

If $(e_i' \supseteq F)$
$P(E) \rightarrow P(E) \cup e_i'$
End do

Partition $P(E) = \bigcup_{j}^{d} PS_j$ //d is the number of distinct partitions

For each $e_i \in E$, $\partial_i \in PS_j$
  if $(ci = \partial_{k'})$
   $\forall (\rho_{k'} \subseteq \partial_{k'})$
Frequency++
If (frequency >minimum_support)
 Set of frequent Sequences $E_f = E_f \cup \rho_{k'}$
End do
End Algorithm RSP

*Fig. 5.0 Algorithm pseudocode*

## IV. RESULTS AND DISCUSSION

Present algorithms for mining sequential patterns are of two categories one that incorporate time window constraint and the others which are generic method without any constraint of event interval. However, incoporation of time constraint is important into mining event-patterns. This improves the quality of knowledge mined from the system. This motivates a need to involve a human expert who can iteratively incorporate his/her learning of the application domain under study into the study of patterns. The additional motivation is many researchers who gave generic framework for mining sequential patterns extended their study to incorporate time window constraint. For example, Agrawal and Srikant [1] have improved their apriori-based method in [1] with an improved algorithm that is apriori based and incorporates maximum time gap, minimum time gap and sliding window constriant [3]. The other perspective of sequential pattern mining is the computational logic of frequent patterns.There are mainly two perspectives to address the problem; one approach is to generate candidates that can be frequent and finding them in the dataset under study and the other is breaking the event elements into parts and accumulating their respective count.Our method Rough Set Partitioning can be compared with GSP [3] proposed by Agrawal and Srikant. As, it incorporate the time window constraint in the mining task. It is found by experiments that our method is atleast 10 times faster than GSP.

The RSP algorithm arranges the data in lexicographical order in each partition which is similar to set enumeration tree as is created in SPAM [4], which is the most efficient methods among the class of methods for sequential pattern mining. However, the difference is: SPAM's method of frequency accumulation is similar to apriori and it evaluates both frequent itemsets and frequent sequences in the same tree, while the proposed method breaks the problem into parts. While RSP works on pruned search space and utilize the apriori property. SPAM is a candidate-generate and test stategy. Many candidates tested by SPAM do not have even a single occurance in the database under study. Moreover, the method does not incporporate time constraint.

In comparison with GSP, RSP is a logical superior technique due to the following:

*Number of database scans:*
*GSP***:** Multiple scans of database
*RSP*: 1 or 2 depending on the size of database also reads all data into datastructure and computational logic is applied on the same
*Search Space:*
GSP: Full database (large)
RSP: Pruned and Partitioned database
*Search Method:*
*GSP:* Bottom up search technique based on apriori principle. All frequent 1 length events are candiates for frequent length 2 events. IT involves multiple scans and accumulates counts of elements from scratch. It addresses both itemset mining and sequence mining simultaneously.

*RSP:* Dynamic counting of frequencies, It breaks the problem into parts and derives benefit for pruning by LEMMA 1.Also it is strictly based on the elements which have atleast one occurance in the database under study.
*Time Constraint adjustmen:*
*GSP:* Mining task needs to be repeated every time the time window under consideration is adjusted.
*RSP:* Dynamic adjustment: As previsualization of patterns is availible prior to execution of mining task.

We have implemented both RSP and GSP in java- JDK1.3 The algorithm uses the java database connectivity interface to the back end used is MSSQL Server 2005.The logic of iteractive time adjustment and creation of input database as in TABLE II is build using PLSQL and the derived data is fetched into the data structures using jdbc. The machine used is HP Proliant DL580G5 with Intel Xeon CPU 1.6 GHZ processor with 8 GB RAM. The operating system is Ms Windows Server 2003 R2. We have used java's core strength of reuse and harnessed the strength of search methods on java's data structures Treeset and ArrayList. The search methods in the above data structures give rapid results in log (n) time where n is the number of elements. Thus finding all elements in database of size N will take Nlog (n) computational time to build treeset of frequent itemsets.

We have tested the efficiency of our method on real and synthetic datasets. The synthetic datasets are same as obtained by using the synthetic data generation program availible at http://www.almaden.ibm.com/cs/quest. The following are the descriptions of the parameters of the dataset.

|D|    size of the database (number of customers)
|C|    Average number of transactions per customer
|I|    Average size of itemset in maximal potentially large
        sequence
|N|    Number of items

On synthetic datasets our algorithm was about 10 times faster than GSP. The runtime comparison of GSP (naive method) with Rough Sets partitioning on synthetic dataset C15-I1-N15-D400 is as in Figure 6.0 The graphs clearly indicated RSP outperforms GSP by over 10 times in execution times. Also as minimum support increases RSP runs even faster due to pruning of more numbers of infrequent itemsets due to Lemma 1 and Lemma 2. RSP gives good performance even on large datasets when the maximal frequent sequence is not very long. Due to logic built on subsequence evaluation, its perfomance degrades on very long sequences but is still better than GSP. It is very efficient as compared to GSP even with large datasets. The runtime evaluation on two more synthetic datasets is given in Figure 7.0 and Figure 8.0

We have also implemented our algorithm on real data of network faults in telecom landline network of madhya pradesh state in India. The data comprised of 47640 records with voice related gross faults collected over a time window of three months. There are 215 distinct elements in the sequence and maximum length of the sequence is 15. The time period of 3 months is adjusted by the analyst by using the previsualization tool proposed in the article. The data comprises of gross faults that have occurred at a landline location in the state of Madhya Pradesh in INDIA. The results of the method helps in finding sequential patterns of the form "The line disturbance due to joint fault in SP-DP section results in a Dead Phone" Thus line disturbance due to joint fault in SP-DP section is the antecedent event and dead phone is the subsequent event. The results help in proactive maintenance of network health and cuts down the revenue loss to network downtime. The algorithm can serve as a decision support system for the organization.

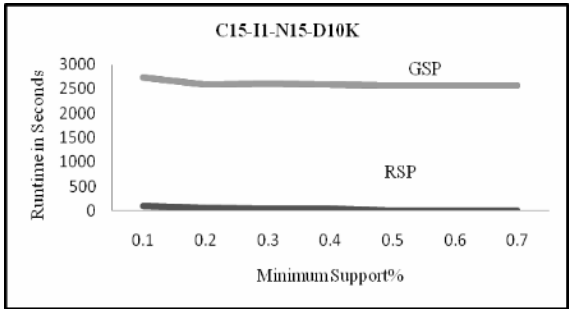

*Fig. 6.0 Runtime comparisons on C15-I1-N15-D400*

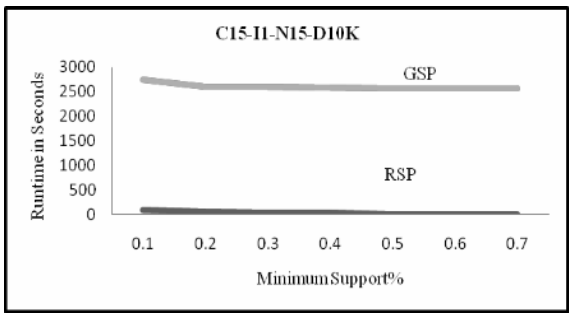

*Fig. 7.0 Runtime Comparisons on C15-I1-N15-D10k*

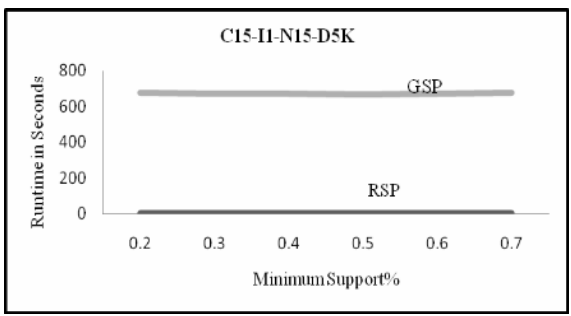

*Fig. 8.0 Runtime comparisons on C15-I1-N15-D5k*

V. CONCLUSION

The following are the benefits of proposed model:

(i) Since support counting is usually the most costly step in sequential pattern mining, proposed technique improves the performance greatly by avoiding costly scanning. The use of the heuristic that all frequent sequences are supersets of frequent item sets largely prunes the search space in terms of reduction in frequent candidate sequences. Also the algorithm is strictly based on elements that exist in the database inder study. In addition, time constraint imposition is seperated from the algorithm and is checked at the database level. The partitions once constructed and stored can be used to mine further data increments in the database.

(ii) It is assumed that all data can be fetched into main memory in a single scan of the database. This assumption is realistic as size of main memory is expanding in giga bytes; also advances in technology to harness the power of using main memories of parallel nodes in a distributed computing environment support the assumption. In addition, the same assumption is used in [13].

(iii) The creation of equivalence classes by indiscernibility relation greatly reduces the search time as it partitions the large datasets into smaller fragments and search methods generate results in smaller time as compared to searching through the whole search space.

(iv) The dynamic frequency accumulation sceme in each partition saves computaiton time.

(v) While other methods search the whole search space, our method partitions the problem into subproblems.

(vi) Lemma 1 prunes the search space. It reduces the length of possible maximal sequence resulting in complexity reduction through Lemma 2.

(vii) Based on experimental results obtained and depicted in graphs, we conclude that RSP is atleast 10 times faster than GSP.

ACKNOWLEDGMENT (HEADING 5)

We are grateful to Department of Biotechnogolgy, New Delhi and MP Council of Science and Technology for providing infrastucture facility for MANIT Bhopal for carrying out this research work. We are also grateful to BHARTI AIRTEL LTD. (LANDLINE) Bhopal for giving us support for implementing our algorithm.

## AUTHORS PROFILE

Jigyasa Bisaria is a research fellow with the Department of Mathematics Maulana Azad National Institute of Technology.Bhopal India. Her research interests are predictive data mining and its applications to real world problems.

Dr. Namita Srivastava is working as Assistant Professer with the Department of Mathematics, Maulana Azad National Institute of Technology. She obtained her PhD. in Mathemetics in 1992 in crack problem. Her current research interest are data mining and its applications.

Dr. Kamal raj Pardasani is working as Professor and Head with the Department of Mathematics and Dean Research and Development Maulana Azad National Institute of Technology, Bhopal. He did his PhD. in applied Mathematics in 1988.
His current research interests are computational biology, data warehousing and mining, bio-computing and finite element modeling.